\documentclass{icrc}
\usepackage{graphicx}

\begin{document}

\title{Secondary Antiprotons in Cosmic Rays}
\author[1,2]{I. V. Moskalenko}
\affil[1]{NRC-NASA/Goddard Space Flight Center, Code 660, Greenbelt, MD 20771, U.S.A.}
\affil[2]{Institute of Nuclear
   Physics, M.\ V.\ Lomonosov Moscow State University, 119 899 Moscow, Russia}
\author[3]{A. W. Strong}
\affil[3]{Max-Planck-Institut f\"ur extraterrestrische Physik, Postfach 1312, 
   85741 Garching, Germany}
\author[4]{J. F. Ormes}
\affil[4]{NASA Goddard Space Flight Center, Code 660, Greenbelt, MD 20771, U.S.A.}
\author[5]{M. S. Potgieter}
\affil[5]{Unit for Space Research, Potchefstroom University for CHE, 
2520 Potchefstroom, South Africa}
\author[5]{U. W. Langner}

\correspondence{imos@milkyway.gsfc.nasa.gov}

\firstpage{1}
\pubyear{2001}


\maketitle

\def\gray{$\gamma$-ray}
\def\grays{$\gamma$-rays}

\begin{abstract}

High energy collisions of cosmic ray (CR) nuclei with  interstellar
gas are believed to be the mechanism producing  the majority of CR
antiprotons.  The distinguishing spectral shape with  a maximum at 2
GeV and a sharp decrease towards lower energies makes antiprotons a
unique probe of the models of particle propagation in the Galaxy and
modulation in the heliosphere.  Besides, accurate calculation of the
secondary antiproton flux provides a ``background'' for searches for
exotic signals from the annihilation of supersymmetric particles  and
primordial black hole evaporation.  Recently new data with large
statistics on the  antiproton flux have become available which allow
for such tests to be performed.  We use our 3D Galactic cosmic ray
propagation code GALPROP to calculate interstellar propagation in
several models.  For our best model we make predictions of proton
and antiproton fluxes near the Earth for different modulation levels
and polarity using a steady-state drift model for heliospheric
modulation.
\end{abstract}

\section{Introduction}
 
Most of the CR antiprotons observed near the Earth are secondaries
produced in collisions of energetic CR particles with interstellar gas
\citep{mitchell}.  Due to the kinematics of the process, the spectrum
of antiprotons has a unique shape distinguishing it from other
cosmic-ray species: it is expected to peak at about 2 GeV decreasing
sharply towards lower energies.  In addition to secondary antiprotons
there are possible sources of primary antiprotons such as candidate
dark matter particles and evaporating black holes.

Despite numerous efforts and overall agreement on the secondary nature
of the majority of CR antiprotons, published estimates of the expected
flux significantly differ (see e.g.\ Fig.\ 3 in Orito et al. 2000).
The major sources of uncertainties are: (i) incomplete knowledge of
cross sections for antiproton production, annihilation, and
scattering, (ii) parameters and models of particle propagation in the
Galaxy, and (iii) modulation in the heliosphere.  While the
interstellar antiproton flux is affected only by uncertainties in the
cross sections and propagation models, the final comparison with
experiment can only be made after correcting for the solar
modulation. Besides, the spectra of CR nucleons have been directly
measured only inside the heliosphere while we need to know the
spectrum  outside in interstellar space to compute the antiproton
production rate correctly.

We have developed a numerical method and corresponding computer code
GALPROP for the calculation of Galactic CR propagation in 3D
\citep{SM98}. The code has been shown to reproduce simultaneously
observational data of many kinds related to CR origin and propagation
\citep{SM98,MS98,MSR98,SMR00}. The code has been validated on direct
measurements of nuclei, antiprotons, electrons, and positrons, and
astronomical measurements of \grays\ and synchrotron radiation.  These
data provide many independent constraints on model parameters.

Here we use the GALPROP code for accurate calculation of production
and propagation of secondary antiprotons.  We explore the dependence
of the antiproton flux on the  nucleon injection spectrum and
propagation parameters.  The antiproton production is calculated using
the $pp$ production cross section and DTUNUC nuclear factors
\citep{simon98} or the $pp$ production cross section scaled
appropriately with atomic numbers.   Inelastic scattering  producing
``tertiary'' antiprotons  and ``secondary'' protons is taken into
account.    The calculated local interstellar spectrum (LIS) is
modulated using the steady-state drift model.  For the calculation
reported here, we use a cylindrically symmetrical Galactic geometry.

\begin{table*}[bt]
\caption{Propagation parameter sets.
\label{table1}} \small
\begin{tabular}{lcccccc}
\hline\noalign{\smallskip}
                     & 
Injection            &
\multicolumn{2}{c}{Diffusion coefficient$^b$}&
&
\multicolumn{2}{c}{Reacceleration/Convection}\\
\cline{3-4}\cline{6-7}
\noalign{\smallskip}

Model                & 
index$^a$, $\gamma$  &
$D_0$, cm$^2$ s$^{-1}$&
Index, $\delta$      &
&
$v_A$, km s$^{-1}$   &
$dV/dz$, km s$^{-1}$ kpc$^{-1}$\\
\hline \noalign{\smallskip}

Stochastic           
Reacceleration (SR)  &
2.43                 &
$6.10\times10^{28}$  &
0.33                 & 
&
30                   &
--                   \\

Minimal
Reacceleration     
\& Convection (MRC)&
2.43                 &
$4.30\times10^{28}$  & 
0.33                 & 
&
17                   &
10                   \\

Plain Diffusion (PD) &
2.16                 &
$3.10\times10^{28}$  & 
0.6                  &
&
--                   &
--                   \\


Diffusion plus      
Convection (DC)      &
$2.46/2.16^c$      &
$2.50\times10^{28}$  & 
$0./0.6^b$              &
&
--                   &
10                 \\
\hline
\end{tabular}

\footnotesize{
$^a$For a power-law in rigidity, $\propto\rho^{-\gamma}$.\\
$^b$$D=\beta D_0(\rho/4\,{\rm GV})^\delta$, index $\delta$ is shown below/above 4 GV.\\
$^c$Index below/above rigidity 20 GV.
}
\end{table*}

\section{Propagation models and parameters}

The propagation parameters have been fixed using the B/C ratio.
Nucleon injection spectra were chosen to reproduce the  local CR
measurements. The source abundances of all isotopes $Z\leq28$ are
given in \citet{SM01}.  We thus use the same (Webber et al.)
cross-section parametri\-zation  as in that work in our calculations.

In all cases the halo size has been set to $z_h=4$ kpc, which is
within the range $z_h=3-7$ kpc derived using the GALRPOP code and the
combined measurements of radioactive isotope abundances, $^{10}$Be,
$^{26}$Al, $^{36}$Cl, and $^{54}$Mn (Strong and Moskalenko 2001 and
references therein).  Note that the exact value of $z_h$ is
unimportant for antiproton calculations provided that the propagation
parameters are tuned to match the B/C ratio.

The ``tertiary'' antiprotons (inelastically scattered secondaries),
significant at the lowest energies, are important in interstellar
space, but make no difference when compared with measurements in the
heliosphere.

\begin{figure}[!tb]
\includegraphics[width=.48\textwidth]{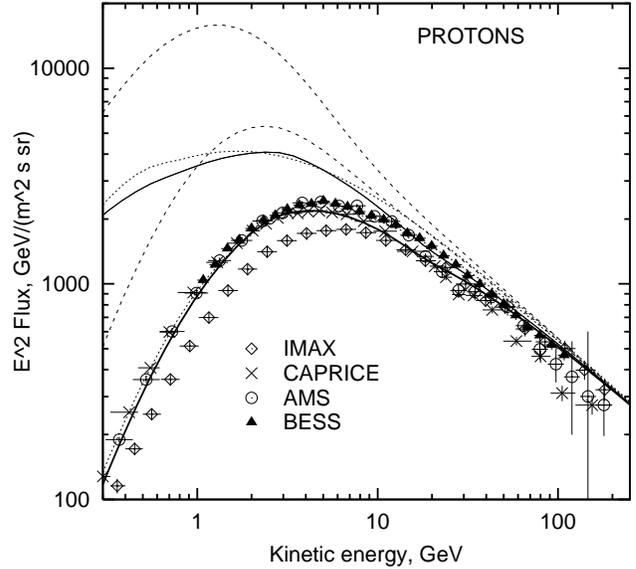}
\caption[fig1.ps]{Calculated proton LIS and 
modulated spectra ($\Phi=550$ MV). The
line coding: solid lines -- DC model (thick line -- modulated
spectrum),  dashes -- DR, dots -- PD. The upper curve is always LIS
spectrum, the lower is modulated.
Data: IMAX \citep{Menn00}, CAPRICE \citep{Boez99}, 
AMS \citep{p_ams}, BESS \citep{Sanu00}.}
\label{fig:protons}
\end{figure}

\begin{figure}[!tb]
\includegraphics[width=.48\textwidth]{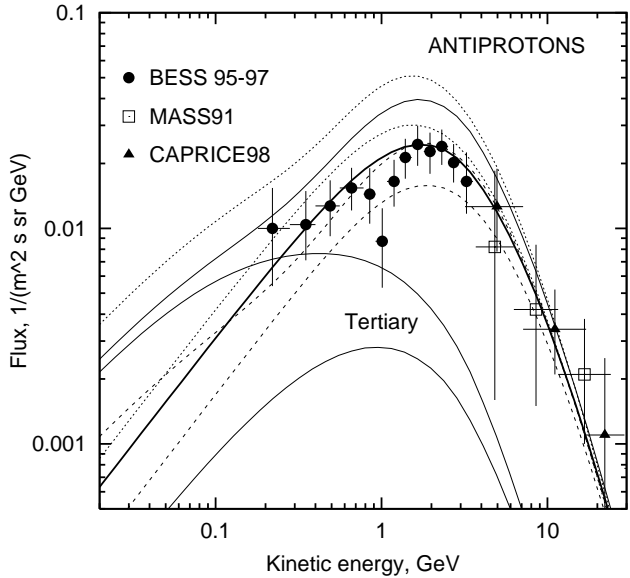}
\caption[fig2.ps]{Calculated spectrum of secondary $\bar p$'s for 3 models.
For each model, the upper curve is the LIS, 
the lower curve -- modulated ($\Phi=550$ MV). 
Line coding as in Fig.\ \ref{fig:protons}.
The two lowest curves show separately the
LIS and modulated spectrum of the ``tertiary'' component in DC model. 
Data: BESS \citep{Orito00}, MASS91 \citep{HE_pbar}, CAPRICE98 \citep{boezio01}.}
\label{fig:pbars}
\end{figure}

\begin{figure}[!tb]
\includegraphics[width=.48\textwidth]{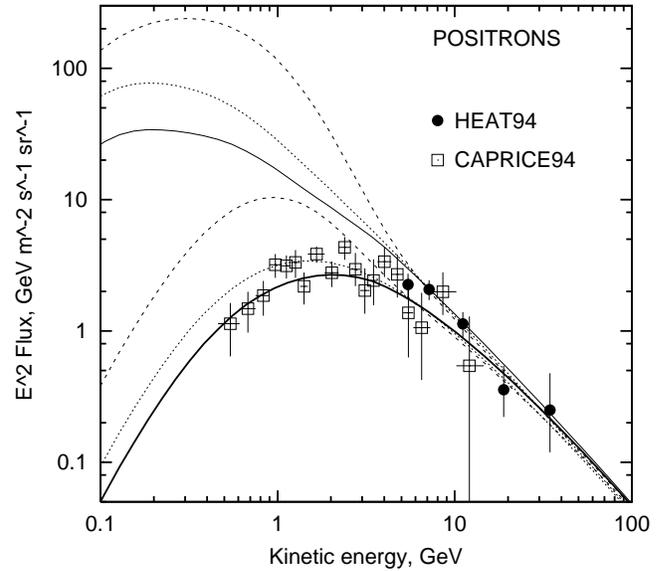}
\caption[fig3.ps]{Calculated positron flux for 3 models.
For each model, the upper curve is the LIS, 
the lower curve -- modulated ($\Phi=600$ MV).
Line coding as in Fig.\ \ref{fig:protons}.
Data: HEAT94 \citep{barwick98}, CAPRICE94 \citep{boezio00}.}
\label{fig:pos}
\end{figure}

To investigate the range of interstellar spectra and propagation
parameters we  considered four basic models (Table \ref{table1}).  Our
results are plotted in Figs.\ \ref{fig:protons}-\ref{fig:pos}.

A model with the stochastic reacceleration (SR) reproduces the sharp
peak in secondary to primary  nuclei ratios in a physically
understandable way without breaks in the diffusion  coefficient and/or
the injection spectrum \citep{SM98,SM01}. However, this model produces
a bump in proton and He spectra at $\sim2$  GeV/nucleon which is not
observed.\footnote{A similar bump is produced also in the electron
spectrum.} This bump can be removed by choosing an injection spectrum
that hardens at  low energies \citep{jones01}.  There are however some
problems with secondaries such as positrons and antiprotons that are
more difficult to manage.  A similar bump appears in the positron
spectrum at $\sim1$ GeV  (Fig.\ \ref{fig:pos}), and the  model
underproduces antiprotons at 2 GeV by more than 30\% (Fig.\
\ref{fig:pbars}).  Taken together they provide evidence against strong
reacceleration\footnote{ We define the reacceleration to be
``strong'' if the model is able to match the B/C ratio without
invoking other mechanisms such as convection and/or breaks in the
diffusion coefficient.} in the ISM.

Another model combining reduced reacceleration and convection (MRC)
also produces too few antiprotons.

Using a plain diffusion model (PD) we can get good agreement with B/C
above few GeV/nucleon, with nucleon spectra and  positrons, but this
model overproduces antiprotons at 2 GeV by $\sim20$\% and contradicts
the secondary/primary nuclei ratio (B/C) below 1 GeV/nucleon.

A model including diffusion and convection (DC) is our best fitting model.  It
reproduces all the particle data ``on average'',  although it has
still some problem with the reproduction of the sharp peak in B/C
ratio. In this model a  flattening of the diffusion coefficient below
4 GV is required to match the B/C ratio at low energies.

To better match primaries ($p$, He) in the DC model, we introduced a
steeper injection spectrum below 20 GV; such a break,  however, has
almost no effect on secondaries ($\bar p$, $e^+$).  The existence of a
sharp upturn below a few GeV/nucleon follows from SNR shock
acceleration theory \citep{ellison}; this is a transition region
between thermal and non-thermal particle populations in the shock. Our
model does not require a sharp break (0.3 in index is enough).

\begin{figure*}[!tb]
\includegraphics[width=.5\textwidth]{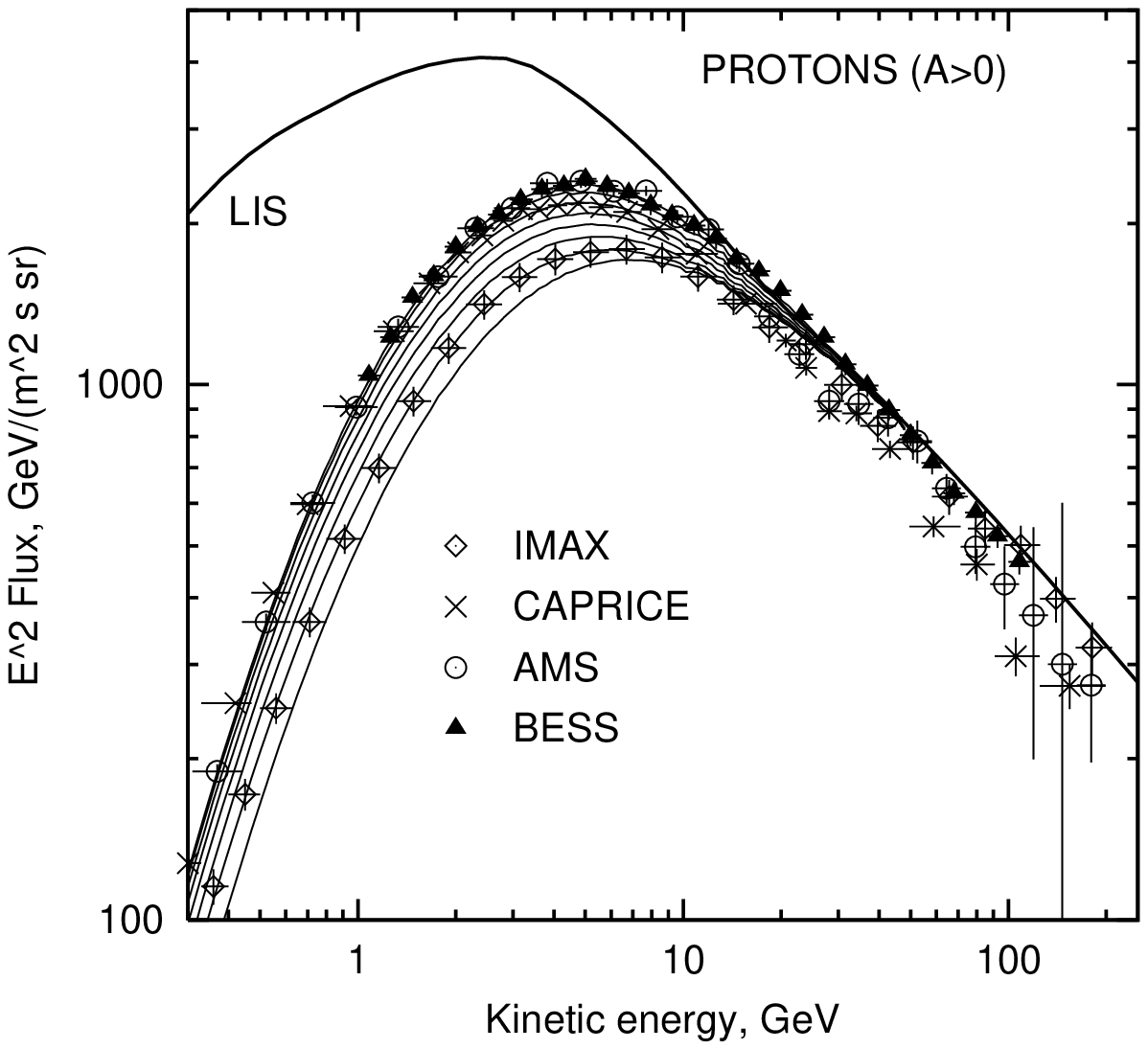}
\includegraphics[width=.5\textwidth]{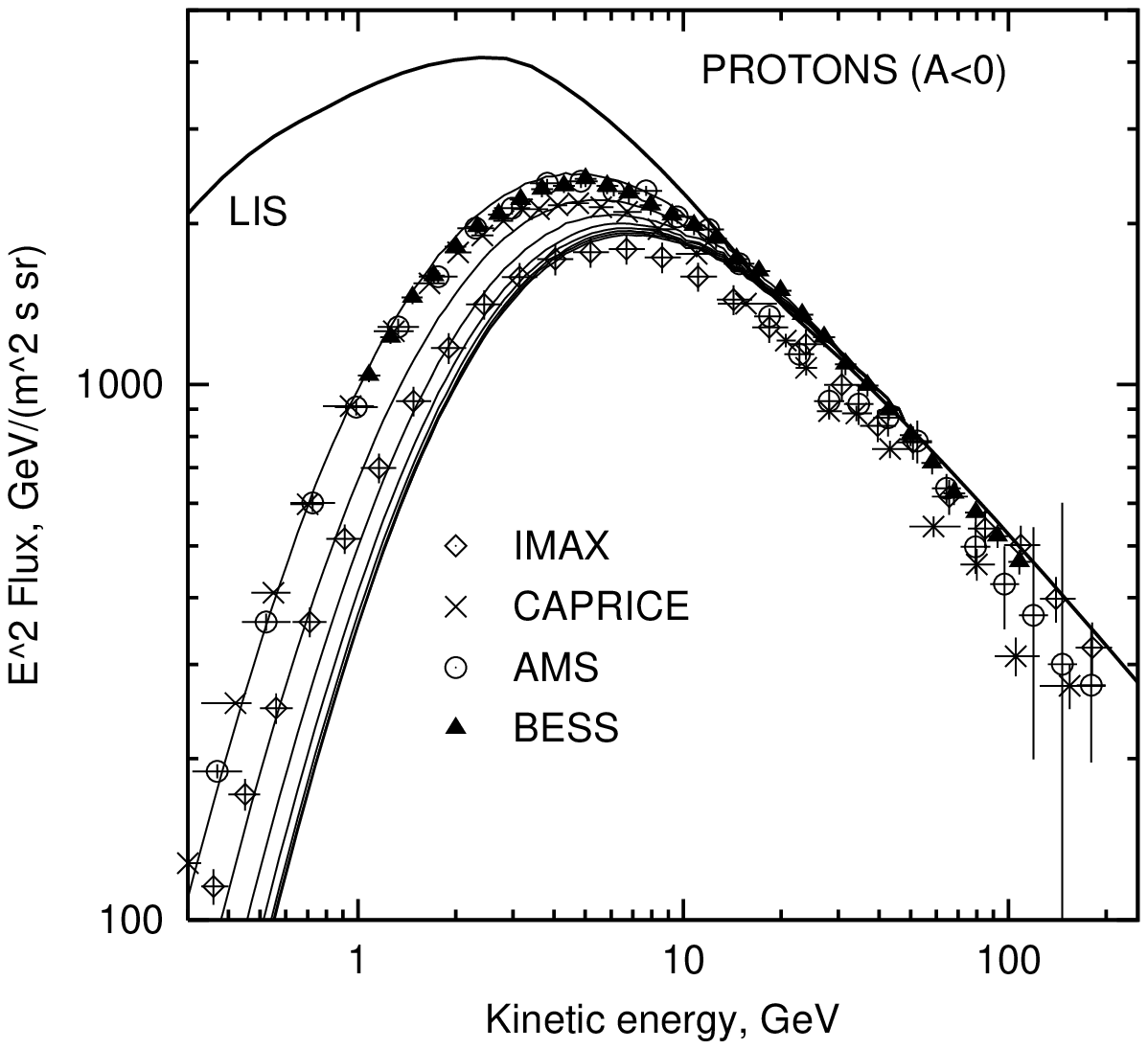}
\caption[fig4a.ps,fig4b.ps]{ 
Calculated proton LIS and 
modulated spectra for the two magnetic polarity dependent
modulation epochs, $A>0$ (left) and $A<0$ (right).
Tilt angle from top to bottom: $5^\circ, 15^\circ,
25^\circ, 35^\circ, 45^\circ, 55^\circ, 65^\circ, 75^\circ$.  
The tilt angle corresponding
to BESS and AMS data is $\sim\!5^\circ\!-15^\circ$ ($A>0$) 
depending on the coronal field model.
On the right panel, $A<0$, the data are shown only for guidance.
Data refences as in Fig.\ \ref{fig:protons}.}
\label{fig:prot}
\end{figure*}
\begin{figure*}[!tb]
\includegraphics[width=.5\textwidth]{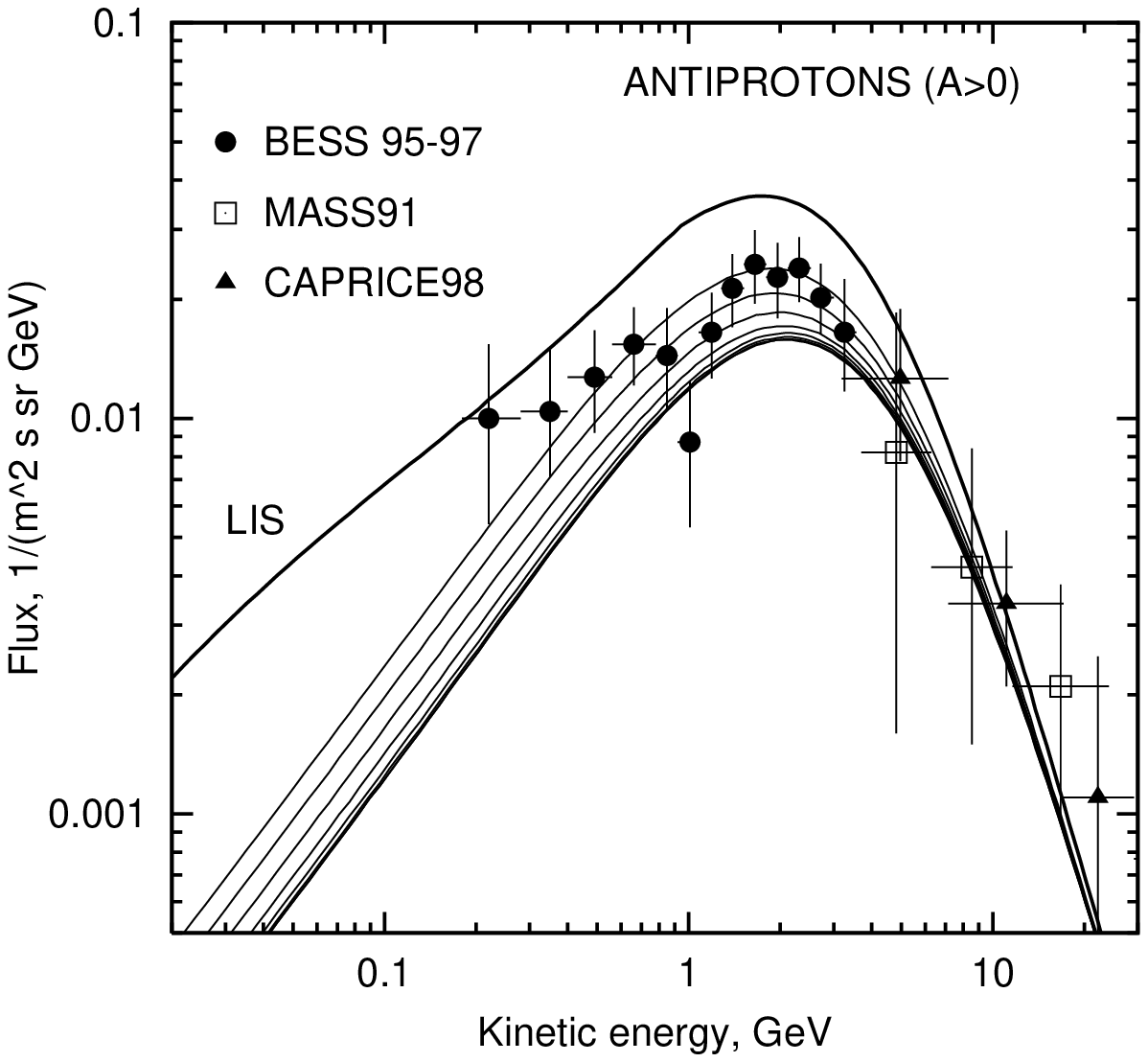}
\includegraphics[width=.5\textwidth]{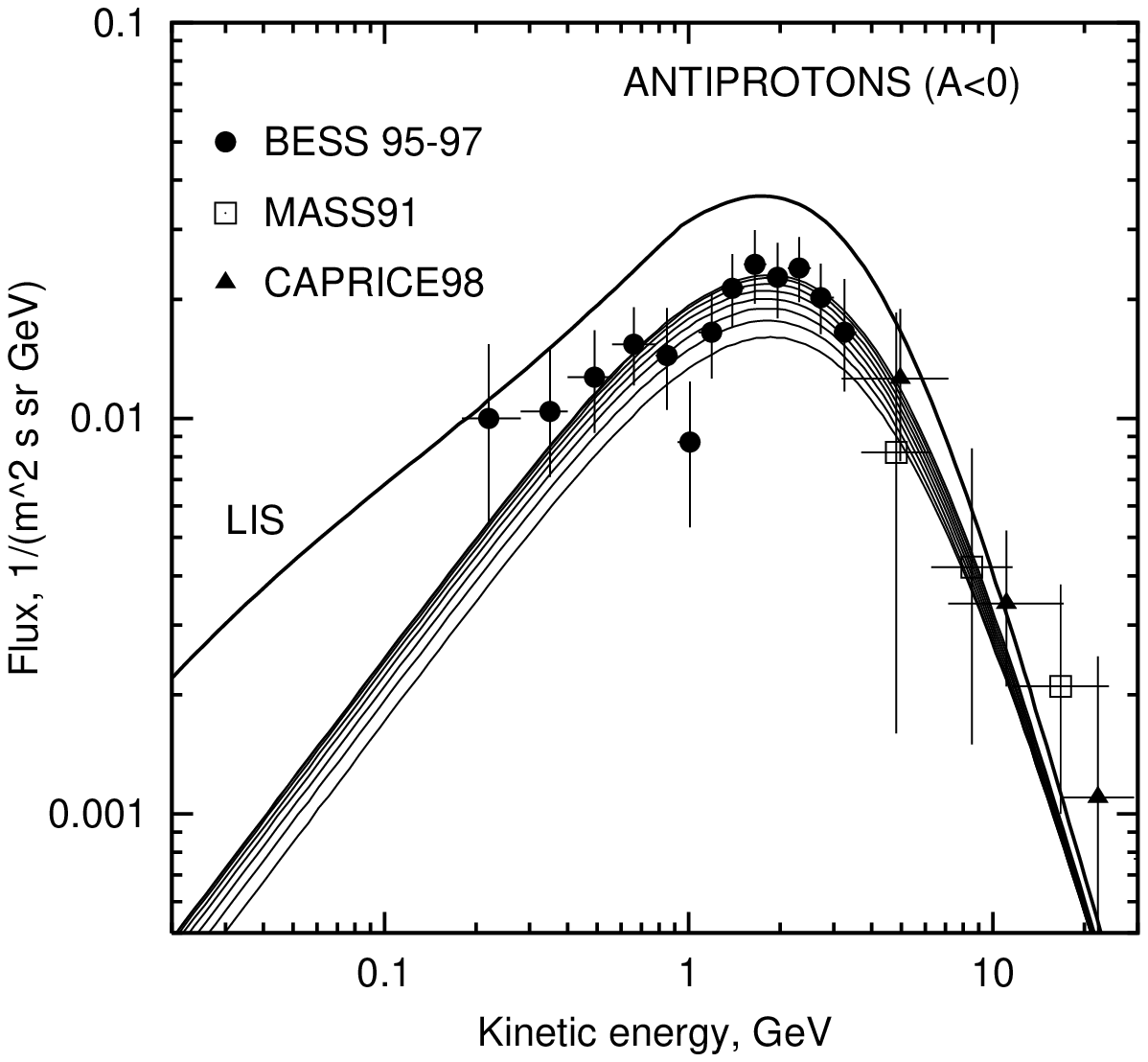}
\caption[fig5a.ps,fig5b.ps]{
Calculated antiproton LIS and 
modulated spectra for the two magnetic polarity dependent
modulation epochs, $A>0$ (left) and $A<0$ (right).
Tilt angle from top to bottom: $5^\circ, 15^\circ,
25^\circ, 35^\circ, 45^\circ, 55^\circ, 65^\circ, 75^\circ$.
The tilt angle corresponding
to BESS data is $\sim5^\circ-15^\circ$ ($A>0$) depending on the coronal field model.
On the right panel, $A\!<\!0$, the data are shown only for guidance.
Data references as in Fig.\ \ref{fig:pbars}.} 
\label{fig:pbar}
\end{figure*}

We use the DC model to calculate the LIS spectra of protons and
antiprotons and then use the drift model to determine their modulated
spectra and ratio over the solar cycles with positive ($A>0$) and
negative ($A<0$) polarity (Figs.\ \ref{fig:prot}, \ref{fig:pbar}).
The variations shown depend on the tilt angle.\footnote{ For Hoeksema's
tilt angle models and updates see
http://quake.stanford.edu/$\sim$wso/}   This allows us to estimate
the near-Earth spectra for arbitrary epochs in the past as well as
make some predictions for the future. It may be also used to test the
theory of heliospheric modulation \citep{potgieter}.

\section{Discussion}

It appears quite difficult to get agreement with B/C, p \& He spectra,
and antiprotons spectrum simultaneously in the framework of simple ``physical'' 
reacceleration/convection \linebreak[4] models; this conclusion is mainly the 
result of the increased precision
of the CR experimental data, and also the improved reliability of the
calculations.

What could be the origin of this failure, apart from the propagation
models ?  For B/C, it seems unlikely that the  cross-sections are
sigificantly in error since for B production from C and O they are
well measured at the relevant energies. For $p$ and He the GeV
bump produced by reacceleration seems to be outside the limits allowed
by the modulation models and the observed fluxes, although the
uncertainties in modulation are still considerable. Such a 
bump in the spectra of primaries ($p$, He, $e^-$) could be ``removed'' 
by including a flattening of the injection spectrum at low energies, while 
for secondaries ($e^+$, $\bar p$) we have no such freedom and the problem remains.

If we assume then that the problem is in the propagation models, we have
shown that it \emph{is} possible to construct a model (DC) which fits
all these data, by postulating a significant flattening of the
diffusion coefficient below 4 GV together with convection 
(and possibly with reduced reacceleration).  
This type of break in the diffusion coefficient is
reminiscent of the standard procedure in ``leaky-box'' models where the
escape time is set to a constant below a few GeV. This has always
appeared a completely \emph{ad hoc} device without physical
justification, but the present analysis suggest it may be forced on us
by these type of  data, so that possibilities for its physical origin
should be studied.  At the same time the sensitivity of the whole
analysis to the modulation models has to be investigated further.

\begin{acknowledgements}
We thank Frank Jones and Vladimir Ptuskin for fruitful 
discussions, and Don Ellison for discussions and a copy of
his manuscript prior to publication.
IVM acknowledges support from NAS/NRC Research 
Associateship Program.
\end{acknowledgements}

\end{document}